%% file: 0Paper.tex
\documentclass[sigconf]{acmart}

\renewcommand\footnotetextcopyrightpermission[1]{} 
\AtBeginDocument{%
  \providecommand\BibTeX{{%
    \normalfont B\kern-0.5em{\scshape i\kern-0.25em b}\kern-0.8em\TeX}}}

\usepackage{soul} 
\usepackage{graphicx}
\usepackage{booktabs}
\usepackage{float}
\usepackage{overpic}
\usepackage{float}  
\usepackage{subfloat}
\usepackage{stfloats} 
\usepackage[skip=0.5\baselineskip]{caption}
\usepackage{bm}
\usepackage{amsmath}
\usepackage{booktabs}
\usepackage{multirow}
\usepackage{multicol}
\usepackage{enumitem}
\usepackage{subcaption}
\usepackage{threeparttable}
\usepackage{xspace}
\usepackage{color}
\usepackage[normalem]{ulem}
\usepackage{algorithmicx,algorithm}
\usepackage{algpseudocode}
\usepackage{algorithm,algpseudocode}
\usepackage{marvosym}   
\usepackage{colortbl}   
\usepackage{tcolorbox}  
\tcbuselibrary{breakable}   
\usepackage{wasysym}    
\usepackage[figuresright]{rotating} 
\usepackage{pifont} 

\makeatletter
\def\@ACM@checkaffil{
    \if@ACM@instpresent\else
    \ClassWarningNoLine{\@classname}{No institution present for an affiliation}%
    \fi
    \if@ACM@citypresent\else
    \ClassWarningNoLine{\@classname}{No city present for an affiliation}%
    \fi
    \if@ACM@countrypresent\else
        \ClassWarningNoLine{\@classname}{No country present for an affiliation}%
    \fi
}
\makeatother

\newcommand{\modelname}{\textsc{R$^2$led}}

\usepackage{xspace}
\newcommand{\etal}{\emph{et al.}\xspace}
\newcommand{\eg}{\emph{e.g.,}\xspace}
\newcommand{\ie}{\emph{i.e.,}\xspace}
\newcommand{\etc}{\emph{etc.}\xspace}

\copyrightyear{2026}
\acmYear{2026}
\setcopyright{acmcopyright}
\acmDOI{XXXXXXX.XXXXXXX}
\acmISBN{978-1-4503-8332-5/26/08}
\begin{document}
\begin{sloppypar}   
\title{\modelname: Equipping Retrieval and Refinement in Lifelong User Modeling with Semantic IDs for CTR Prediction}

\author{Qidong Liu$^{1}$\dag, Gengnan Wang$^2$\dag, Zhichen Liu$^3$, Moranxin Wang$^1$, Zijian Zhang$^{4}$, Xiao Han$^{5}$ \\
Ni Zhang$^{1}$, Tao Qin$^{1}$, Chen Li$^{1}$}
\affiliation{
    \institution{$^1$Xi'an Jiaotong University, 
    $^2$University of Edinburgh, $^3$Nanyang Technological University, \\
    $^4$Jilin University, 
    $^5$Zhejiang University of Technology}
    \country{}
}
\email{{liuqidong, nizhang, cli}@xjtu.edu.cn, s2795617@ed.ac.uk, ZHICHEN001@e.ntu.edu.sg}
\email{wangmo@stu.xjtu.edu.cn, zhangzijian@jlu.edu.cn, hahahenha@gmail.com, qin.tao@mail.xjtu.edu.cn}
\thanks{\dag~ Both authors contributed to this paper equally.}

\renewcommand{\shortauthors}{Qidong Liu and Gengnan Wang \etal}
\renewcommand{\shorttitle}{Equipping Retrieval and Refinement in Lifelong User Modeling with Semantic IDs for CTR Prediction}

\begin{abstract}
  Lifelong user modeling, which leverages users' long-term behavior sequences for CTR prediction, has been widely applied in personalized services. Existing methods generally adopted a two-stage ``retrieval-refinement'' strategy to balance effectiveness and efficiency. 
  However, they still suffer from (i) noisy retrieval due to skewed data distribution and (ii) lack of semantic understanding in refinement. While semantic enhancement, \eg LLMs modeling or semantic embeddings, offers potential solutions to these two challenges, these approaches face impractical inference costs or insufficient representation granularity.
  Obsorbing multi-granularity and lightness merits of semantic identity (SID), we propose a novel paradigm that equips \textbf{\underline{R}}etrieval and \textbf{\underline{R}}efinement in \textbf{\underline{L}}ifelong User Modeling with S\textbf{\underline{E}}mantic I\textbf{\underline{D}}s (\textbf{\modelname}) to address these issues.
  First, we introduce a Multi-route Mixed Retrieval for the retrieval stage. On the one hand, it captures users' interests from various granularities by several parallel recall routes. On the other hand, a mixed retrieval mechanism is proposed to efficiently retrieve candidates from both collaborative and semantic views, reducing noise. Then, for refinement, we design a Bi-level Fusion Refinement, including a target-aware cross-attention for route-level fusion and a gate mechanism for SID-level fusion. 
  It can bridge the gap between semantic and collaborative spaces, exerting the merits of SID. 
  The comprehensive experimental results on two public datasets demonstrate the superiority of our method in both performance and efficiency.
  To facilitate the reproduction, we have released the code online\footnote{https://github.com/abananbao/R2LED}.
\end{abstract}



\maketitle

\input{1Introduction}

\input{2Preliminary}

\input{3Method}

\input{4Experiment}

\input{5RelatedWork}
\input{6Conclusion}


\bibliographystyle{ACM-Reference-Format}
\bibliography{main}


\end{sloppypar}
\end{document}

%% file: 1Introduction.tex
\section{Introduction}

\begin{figure}[!t]
\centering
\includegraphics[width=1\linewidth]{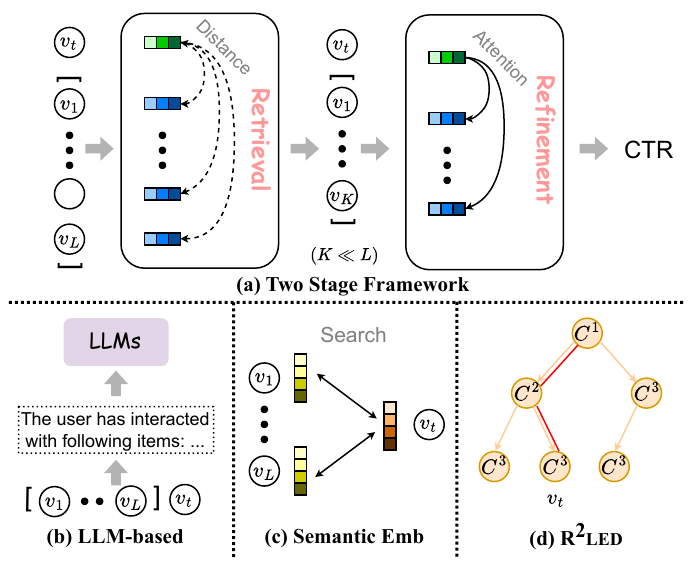}
\caption{The illustrations to the general two-stage strategy and semantic enhancing patterns for lifelong user behavior modeling. (a) The general ``retrieval-refinement'' framework. (b) LLM-based modeling. (c) Semantic Embeddings. (d) Semantic Tree-based Retrieval in \modelname~ (Ours). }
\label{fig:preliminary}
\vspace{-2mm}
\end{figure}

Modern online personalized platforms, such as Kuaishou, Taobao, \etc, have cumulated a large corpus of user behavior records. 
Recent research studies~\cite{xu2025mirrn,chang2023twin,si2024twinv2} have verified the efficacy of leveraging users' lifelong behaviors, \ie long-term historical interactions, to enhance Click-Through Rate (CTR) prediction. 
While pioneering methods like DIN~\cite{zhou2018din} utilize attention mechanisms to capture users' long-term interests, they are often constrained by the high computational costs associated with extremely long interaction sequences. 
To address this, a two-stage ``retrieval-refinement'' strategy, as shown in Figure~\ref{fig:preliminary}(a), has been proposed. 
Typically, these methods~\cite{pi2020sim,chen2021eta} first \textbf{retrieve} target-relevant interactions, and then \textbf{refine} the user's preferences based on the shortened behavior sequences to avoid excessive latency.

The two-stage method achieves dual advantages: it enhances computational efficiency by shortening long sequences and improves prediction effectiveness by eliminating irrelevant historical behaviors. 
However, two significant challenges remain: 
\ding{182} \textbf{Noisy Sensitivity in Retrieval}. Recent advancements have focused on optimizing the retrieval stage for both latency and precision. For example, ETA~\cite{chen2021eta} proposes locality-sensitive hashing (LSH) to reduce the time complexity of searching, while TWIN~\cite{chang2023twin,si2024twinv2} derives a target attention for retrieval to harmonize the two stages. Despite these improvements, they primarily rely on co-occurrence-based item embeddings. Such embeddings are often skewed by the interaction distribution~\cite{lee2024post}, which often introduces noise into retrieval results.
\ding{183} \textbf{Semantics Deficiency in Refinement}. Prevailing research studies~\cite{zhou2018din,xu2025mirrn} capture users' preferences only by collaborative signals, regarding features and interactions as discrete identities. This reliance on shallow ID-based representation ignores rich semantic information in user behaviors, ultimately compromising the performance of models, especially for long-tail items~\cite{liu2024llm}.

Intuitively, integrating semantic information into the two stages of lifelong user modeling could mitigate the aforementioned challenges. 
Following this intuition, two typical lines of work have emerged, though both face significant limitations.
(i) \textbf{LLM-based Modeling} (Figure~\ref{fig:preliminary}(b)). Building on the advancements of large language models (LLMs)~\cite{wu2024survey,zhao2023survey}, several pioneering works~\cite{lin2024rella,zhu2024reclora} fine-tune LLMs for lifelong user modeling. Although these models excel at capturing deep semantics, their high inference costs make them impractical for online services.
(ii) \textbf{Semantic Embeddings} (Figure~\ref{fig:preliminary}(c)). Recent research~\cite{jia2025learn,zhang2024notellm} has revealed the potential of semantic embedding in item-item matching. However, the embedding-based strategy is hindered by insufficient representative abilities of a single dense vector~\cite{liu2025best}. Additionally, semantic embeddings generally require a high-dimensional space, resulting in high latency during the process of large-scale embedding retrieval.

Semantic identity (SID)~\cite{rajput2023recommender,hou2025towards} emerged as a compelling alternative, consisting of a set of discrete codes derived from the quantization of decomposed, multi-granularity semantic embeddings. Given its hierarchical nature and computational efficiency, SID is uniquely positioned to bridge the gap between the two aforementioned semantic strategies.
Nevertheless, integrating SID into the retrieval and refinement paradigm presents two challenges:
(i) \textbf{Retrieval Complexity}. For the retrieval stage, \textit{due to its multi-levels, traditional matching mechanisms struggle to efficiently navigate the resulting search space, complicating the item-retrieval process.
} (ii) \textbf{Space Misalignment}. For the refinement stage, general features are modeled in collaborative space, while SID is originally in semantic space. Besides, the interest spaces in different granularities also vary. \textit{These misalignments prevent the refinement stage from effectively fusing these signals, leading to suboptimal performance.
}


To effectively mitigate the challenges of \ding{182} Noisy Sensitivity and \ding{183} Semantic Deficiency, we propose a novel framework, equipping \textbf{\underline{R}}etrieval and \textbf{\underline{R}}efinement in \textbf{\underline{L}}ifelong User Modeling with S\textbf{\underline{E}}mantic I\textbf{\underline{D}}s (\textbf{\modelname}).
In the retrieval stage, we introduce a Multi-route Mixed Retrieval (MMR) mechanism. Among the MMR, several parallel retrieval routes are designed to capture the user's interest at various granularities. Besides, we propose a mixed retrieval strategy to aggregate candidates from both collaborative and semantic perspectives, reducing the impact of interaction noise. 
Specifically, a SID-tree-based strategy is derived to alleviate the retrieval complexity.
In the refinement stage, SIDs are integrated as enriched semantic features to complement traditional ID-based modeling. Then, we propose a bi-level fusion, which enables fine-grained alignment at route and SID levels.
The contributions are as follows: 
\vspace{-5mm}
\begin{itemize}[leftmargin=*]
    \item We propose an SID-based paradigm, named \modelname, to integrate Semantic Identities into the retrieval-refinement pipeline for lifelong user behavior modeling. 
    
    \item We develop a multi-route retrieval method for SID-based item matching in various granularities and a bi-level fusion refinement to address the problem of space misalignment.
    
    \item Extensive experiments on two public datasets have validated that \modelname~significantly outperforms state-of-the-art methods in both prediction accuracy and inference efficiency.
\end{itemize}

%% file: 2Preliminary.tex
\section{Preliminary}

\subsection{Lifelong User Modeling} \label{sec:pre_lifelong}

In this section, we formulate the CTR prediction task with a specific focus on lifelong user behavior modeling.
Let $\mathcal{U}=\{u_1, u_2, \dots, u_{|\mathcal{U}|}\}$ and $\mathcal{V}=\{v_1, v_2, \dots, v_{|\mathcal{V}|}\}$ denote the set of users and items, where $|\mathcal{U}|$ and $|\mathcal{V}|$ represent the number of users and items. For a specific user $u \in \mathcal{U}$, the historical behavior sequence is denoted as $\mathcal{S}_u = \{v_1, \dots v_i, \dots, v_L\}$, where $v_i \in \mathcal{V}$ represents the $i$-th interacted item and $L$ is the sequence length. Given a user $u$ and a target item $v_t$, the input instance is constructed as $\mathcal{x} = \{u, v_t, \mathcal{S}_u\}$. 
The goal of the CTR task is to learn a prediction model $\hat{y} = f(\mathcal{x})$ to estimate the probability of the user clicking on the target item $v_t$.

With ultra-long behavior sequences, lifelong user modeling is typically formulated as a two-stage framework~\cite{pi2020sim}. In \textbf{Stage-1 (Retrieval)}, a size-controlled Top-$K$ subsequence $\mathcal{M}_{u,v_t}$  is selected from $\mathcal{S}_u$ using query $q$ (refer to target item in general), where $K$ is the pre-defined length of the subsequence: 
\begin{equation}
\begin{aligned}
\mathcal{M}_{u,v_t}=\textsc{Retrieve}(q,\mathcal{S}_u;K),\quad K\ll L.
\end{aligned}
\end{equation}
where $\textsc{Retrieve}(\cdot)$ is a ranking function to get the most similar item set with size of $K$. 
In \textbf{Stage-2 (Refinement)}, the subsequence $\mathcal{M}_{u,v_t}$ will replace $\mathcal{S}_u$ to construct a target-aware input for user, \ie $\tilde{\mathcal{x}}=\{u, v_t, \mathcal{M}_{u,v_t}\}$ , and output the click probability $\hat{y} = g(\tilde{\mathcal{x}}
)$.

\subsection{Semantic Identity (SID)} \label{sec:pre_sid}
Given an item set $\mathcal{V}$, each item $v_i \in \mathcal{V}$ is associated with semantic features $s_i$
(\eg caption or description). Using a pre-trained modality encoder (\eg text embedding model) $E(\cdot): {s} \rightarrow \mathbb{R}^{d}$
transforms the semantic features $s_i$ of item $v_i$ into a $d$-dimensional representation
$\mathbf{h}_i \in \mathbb{R}^{d}$,
\begin{equation}
\mathbf{h}_i = E(s_i).
\end{equation}

Given the semantic feature embedding $\mathbf{h}_i$, a tokenizer function, \ie
$\mathrm{Tokenizer}(\cdot): \mathbb{R}^{d} \rightarrow \{1,2,\ldots,D\}^{L}$
maps $\mathbf{h}_i$ to a length-$L$ discrete semantic ID sequence:
\begin{equation}
C_i = \mathrm{Tokenizer}(\mathbf{h}_i)
= \big[c^{1}_{i}, c^{2}_{i}, \ldots, c^{L}_{i}\big],
\end{equation}
where $D$ refers to the cardinality of each ID, also known as the size of the codebook.
In general, the tokenizer can be selected from Residual Mini-Batch K-Means (RK-Means \cite{deng2025onerec}), Residual Vector Quantization (R-VQ \cite{esser2021taming}), and Residual Quantized Variational Autoencoder (RQ-VAE \cite{lee2022autoregressive}).

%% file: 3Method.tex
\section{Method}

\begin{figure*}[t]
\centering
\includegraphics[width=1.0\linewidth]{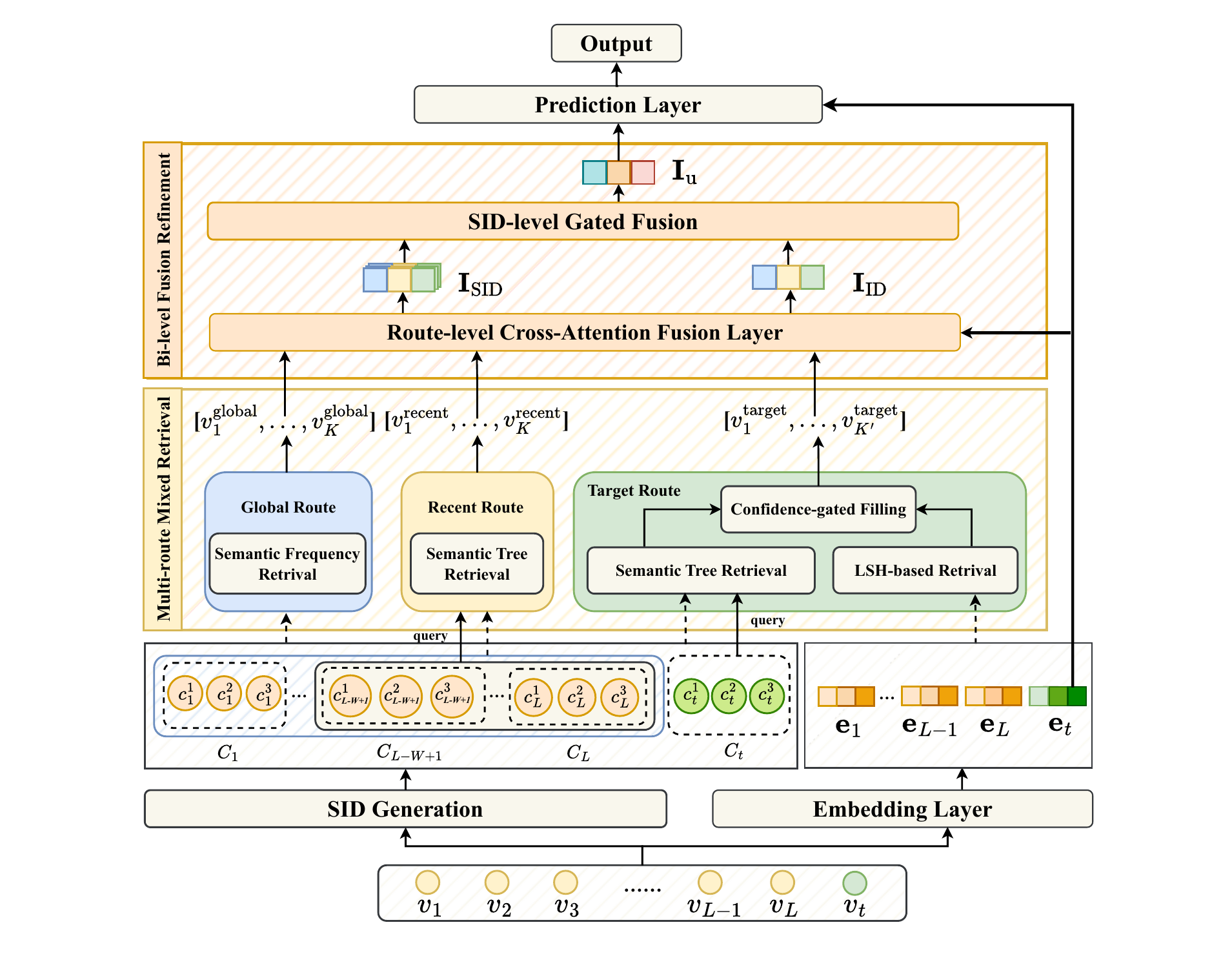}
\caption{The overview of the proposed \modelname.}
\label{fig:framework}
\vspace{-2mm}
\end{figure*}

\subsection{Overview}

In Figure~\ref{fig:framework}, we show the overview of the proposed \modelname, which follows the two-stage strategy mentioned in Section~\ref{sec:pre_lifelong}.
\modelname~is built upon two key components: \textbf{Multi-route Mixed Retrieval (MMR)} for semantic-aware retrieval in \textbf{Stage-1}, and \textbf{Bi-level Fusion Refinement (BFR)} for hierarchical interest fusion in \textbf{Stage-2} to produce a target-aware long-term representation for prediction.

In \textbf{Stage-1}, to mitigate the noise introduced by traditional pure ID-based retrieval in ultra-long user histories, we first assign each item $v_i\in\mathcal{V}$ a tri-level semantic ID tuple $C_i=(c^{1}_i,c^{2}_i,c^{3}_i)$
offline by the SID Generation Module. It organizes semantically similar items into a shared-prefix hierarchy, assisting in retrieving key behaviors that are semantically similar to the target item $v_t$.
Building on this semantic structure, we design a \textbf{Multi-route Mixed Retrieval (MMR)}, illustrated in Section~\ref{sec:method_mmr}, to capture users' preferences in various time granularities. 
MMR will derive multi-route subsequence sets, \ie
$\mathcal{M}_{u,v_t}=\{\mathcal{M}^{T}_{u,v_t},\mathcal{M}^{R}_{u},\mathcal{M}^{G}_{u}\}$
from the ultra-long history $\mathcal{S}_u$ via a \emph{Target} route, a \emph{Recent} route, and a \emph{Global} route, capturing both target-specific preferences and broader semantic interests under noisy histories.
Specifically, we proposed a Semantic Tree Retrieval to recall items efficiently via their SIDs.

In \textbf{Stage-2}, to align target-relevant evidence and integrate ID-based collaborative signals with SID-based semantic signals, we further introduce \textbf{Bi-level Fusion Refinement (BFR)}. It takes the retrieved multi-route Top-$K$ subsequences $\mathcal{M}_{u,v_t}$ as input to model user interests at multiple levels, ranging from exact ID-level signals to tri-level semantic IDs.
Among BFR, we design a target-aware cross-attention to derive route-specific interest representations, \ie $\mathbf{I}_{\rm global}$, $\mathbf{I}_{\rm recent}$ and $\mathbf{I}_{\rm target}$.
Then, these representations are first fused across routes and subsequently fused across semantic levels to obtain the final target-aware representation, detailed in Section~\ref{sec:modelling}.


\subsection{Multi-route Mixed Retrieval (MMR)}
\label{sec:method_mmr}
Considering that retrieving only target-relevant items makes it difficult to capture diverse interests, we construct a \textbf{Multi-route Mixed Retrieval} module based on different time granularities, including \textbf{target}, \textbf{recent}, and \textbf{global} information.


\subsubsection{\textbf{Target Route}}
Information from the target item plays a central role in predicting the user’s click-through rate~\cite{zhou2018din}. Therefore, we construct a target-aware retrieval route by using the target item as the query in semantic and collaborative spaces.

Specifically, we directly use the target item's semantic ID tuple $C_t=(c^{1}_t,c^{2}_t,c^{3}_t)$ as the query $q_t^{\rm SID}$.
Through the designed \textbf{Semantic Tree Retrieval} module, we retrieve the subsequence $\mathcal{M}_{u,v_t}^{\rm target}(\mathrm{SID})$, which consists of historical items most semantically aligned with the target item.
In addition, to enhance the target-aware collaborative information, we adopt a \textbf{LSH-based Retrieval} strategy~\cite{chen2021eta}, where the embedding of the target item $v_t$ is the query $q_t^{\rm ID}$, retrieving the subsequence $\mathcal{M}_{u,v_t}^{\rm target}(\mathrm{ID})$.
Finally, we obtain the target-aware subsequence $\mathcal{M}_{u,v_t}^{\rm target}$ by
\textbf{Confidence-gated Filling}, which fills the LSH-retrieved items
$\mathcal{M}_{u,v_t}^{\rm target}(\mathrm{ID})$ into
$\mathcal{M}_{u,v_t}^{\rm target}(\mathrm{SID})$ according to the discriminability of the LSH-based retrieval.

\vspace{1mm}
\noindent\textbf{Semantic Tree Retrieval}.
In this paper, we propose a tree-based retrieval approach for the target route.
Given the target item $v_t$ and a historical item $v_i$, we compute a strict prefix-based matching score along the SID tree,
where $\mathbb{I}[\cdot]$ denotes the indicator function:
\begin{equation}
\mathbb{I}[x]=
\begin{cases}
1, & \text{if } x \text{ is true},\\
0, & \text{otherwise}.
\end{cases}
\end{equation}

\begin{equation}
\begin{aligned}
\mathrm{Score}(v_i, v_t)
= \sum_{l=1}^{3}  \prod_{j=1}^{l} \mathbb{I}\!\left[c^{j}_{i}=c^{j}_{t}\right].
\end{aligned}
\end{equation}

This scoring rule corresponds to a tree traversal defined by the hierarchical SID: the tuple $C_i=(c^{1}_i,c^{2}_i,c^{3}_i)$ specifies a unique root-to-leaf path in a tri-level prefix tree, and the product term enforces a \emph{strict} prefix constraint.
Specifically, a match at a deeper level contributes to the score only when all higher-level prefixes are matched, \ie only when $v_i$ and $v_t$ share a longer common path in the SID tree.
Therefore, longer matched prefixes naturally indicate finer-grained semantic alignment.

By scoring all behaviors in $\mathcal{A}_u = \{C_{1}, \dots C_{i}, \dots, C_{L}\}$ against the query $q_t^{\rm SID}=C_t$, we obtain the Top-$K$ behaviors with the highest score. And the target-aware subsequence $\mathcal{M}_{u,v_t}^{\rm target}({\rm SID})$ is constructed by selecting historical items obtained via tree-based retrieval with strict prefix matching over hierarchical SIDs:
\begin{equation}
\begin{aligned}
   \mathcal{M}_{u,v_t}^{{\rm target}}({\rm SID})= \textsc{Retrieve}(q_t^{\rm SID}, \mathcal{A}_u; \mathit{K}) 
  = \mathrm{TopK}(\mathrm{Score}(v_i, \mathit{v}_\mathit{t})).
\end{aligned}    
\end{equation}
Here, the $\textsc{Retrieve}(\cdot)$ function takes the semantic behavior sequence $\mathcal{A}_u$ and the query $q_t^{\rm SID}$ as inputs, and outputs a query-aware subsequence $\mathcal{M}_{u,v_t}^{\rm target}(\rm SID)$.

\vspace{1mm}
\noindent\textbf{LSH-based Retrival}.
According to the locality-sensitive property of SimHash~\cite{charikar2002similarity}, we measure the similarity between the ID-based query $q_t^{\bf{ID}}=\mathbf{e}_t$ and historical behaviors $\mathcal{S}_u$ using the Hamming distance between their binary signatures, rather than the inner product in the continuous embedding space.
$\mathbf{e}_i$ and $\mathbf{e}_t$ represent the embeddings of item $i$ and target item, respectively.
We then compute the Hamming distance and rank all historical behaviors by this distance. Finally, we select the Top-$N$ behaviors with the smaller Hamming distances and get subsequence $\mathcal{M}_{u,v_t}^{\rm target}(\rm ID)$:

\begin{equation}
\begin{aligned}
    \mathcal{M}_{u,v_t}^{{\rm target}}({\rm ID}) = \textsc{Retrieve}(q_t^{{\rm ID}}, \mathcal{S}_u; N)
    = \mathrm{TopN}(\mathrm{Hamming}(h_i, h_t)),
\end{aligned}    
\end{equation}
where $h_i = \mathrm{SimHash}(\mathbf{e}_i)$ and $h_t= \mathrm{SimHash}(q_t^{\rm ID})$.

\vspace{1mm}
\noindent\textbf{Confidence-gated Filling}.
To prevent noisy collaborative signals from being injected, we apply a confidence gate to decide whether $\mathcal{M}_{u,v_t}^{\rm target }({\rm ID})$ are used for filling.
The key intuition is \emph{discriminability}: a reliable neighbor set should yield a \emph{distance separation}, \ie truly relevant behaviors are much closer to the target than the remaining candidates.
In contrast, if distances concentrate in a narrow range, the ``nearest'' candidates are not clearly distinguishable, implying low-quality and potentially noisy retrieval.

Following the idea, we first sort the Hamming distances $\{d_i\}_{i \in L}$ in ascending order and denote the smallest distance as $d_{min}$.
To measure how well the remaining candidates are separated from the closest neighbor, we compute the relative distances with respect to $d_{min}$ and normalize them by the hash length $H$:
\begin{equation}
\tilde d_i = \frac{d_i - d_{min}}{H}.
\end{equation}
This normalization removes scale effects and focuses on the \emph{distance gap} between the nearest neighbor and other candidates.
We then adopt the variance of the normalized distances as a quality indicator:
\begin{equation}
\mathrm{Var}_t = \mathrm{Var}\!\left(\{\tilde d_i\}_{i\in L}\right), \qquad
g_t = \mathbb{I}\!\left[\mathrm{Var}_t \ge \tau\right],
\label{eq:fill_gate}
\end{equation}
where $\tau$ is a threshold.
A larger $\mathrm{Var}_t$ indicates that the distances are well spread, meaning that the closest neighbors are clearly separated from less relevant candidates.
Conversely, a smaller variance implies that all candidates lie at similar distances to the target, making the retrieved set less discriminative and potentially noisy.

Finally, we fill $\mathcal{M}_{u,v_t}^{\rm target}(\rm ID)$ into $\mathcal{M}_{u,v_t}^{{\rm target}}(\rm SID)$ based on the confidence gate above, yielding the target-route subsequence $\mathcal{M}_{u,v_t}^{\rm target}$:
\begin{equation}
\begin{aligned}
\mathcal{M}_{u,v_t}^{{\rm target}}=
\mathcal{M}_{u,v_t}^{{\rm target}}({\rm SID})\ \cup\ \big(g_t \cdot \mathcal{M}_{u,v_t}^{{\rm target}}({\rm ID})\big),
\end{aligned}    
\end{equation}
where $g_t \cdot \mathcal{M}$ means keeping the ID-based subsequence only if $g_t=1$. Let $K'$ denotes the size of $\mathcal{M}_{u,v_t}^{{\rm target}}$.

\subsubsection{\textbf{Recent Route}}
Recent behaviors provide direct evidence of a user’s short-term intent and are important for the current click-through decision~\cite{lv2019sdm}.
Therefore, we broaden the query beyond a single target item and construct a \emph{recent semantic window} from the last $W$ behaviors, \ie $\mathcal{R}_u=\{C_{L-W+1},\dots,C_L\}$ as the query $q_{\rm recent}$.
Using this query, we retrieve a subsequence $\mathcal{M}_{u}^{\rm recent}$ that contains the \emph{most semantically relevant} historical items with respect to the recent window, so that it represents the user’s short-term interest.

Notably, we do not directly take $\mathcal{R}_u$ as the retrieved subsequence, since the recent window may include noisy behaviors (\eg accidental clicks).
Instead, we use $\mathcal{R}_u$ only as a semantic descriptor of short-term preference and search the entire long-term sequence for other behaviors that are semantically consistent with short-term interest.
Given the recent window $\mathcal{R}_u$, we treat each recent item as a query and perform tree-based retrieval over the full semantic behavior sequence $\mathcal{A}_u$.
Concretely, for each $v_r \in \{v_{L-W+1},\dots,v_L\}$, we compute a prefix-based tree matching score between $v_r$ and every historical item $v_i$, and obtain a query-specific score ${\rm Score}(v_i, v_r)$.
We then aggregate the scores contributed by all queries in the window to form the final recent-aware relevance score:
\begin{equation}
\begin{aligned}
{\rm Score}_{\rm recent}(v_i)
= \sum_{r=L-W+1}^{L} {\rm Score}(v_i, v_r), \qquad i=1,\dots,L .
\end{aligned}
\end{equation}
Finally, we select the Top-$K$ historical items with the largest ${\rm Score}_{\rm recent}(v_i)$ to construct $\mathcal{M}_{u}^{\rm recent}$:

\begin{equation}
\begin{aligned}
\mathcal{M}_{u}^{\rm recent} &= \textsc{Retrieve}(q_{\rm recent}, \mathcal{A}_u; K)= \mathrm{TopK}({\rm Score}_{\rm recent}(v_i)).
\end{aligned}
\end{equation}

\subsubsection{\textbf{Global Route}}

Recent-route retrieval focuses on short-term behaviors, whereas a user's global preference is often reflected in semantic patterns that recur throughout the entire long-term behavior sequence~\cite{yu2019adaptive}.
Therefore, we construct a global-aware query by considering the full semantic behavior sequence $\mathcal{A}_u$ and retrieve a subsequence  $\mathcal{M}_u^{\rm global}$. This subsequence contains the \emph{globally representative} historical items, so that we can capture the user’s stable long-term interests efficiently.

For each historical item $v_i$, we compute a global frequency relevance score based on how often its hierarchical SID prefixes recur in $\mathcal{A}_u$.
Let $\mathrm{L}_j(v_i)$ denote the length-$j$ prefix of $C_i$, ($\mathrm{L}_1(v_i)=(c^{1}_i)$, $\mathrm{L}_2(v_i)=(c^{1}_i,c^{2}_i)$, $\mathrm{L}_3(v_i)=(c^{1}_i,c^{2}_i,c^{3}_i)$).
We define the global score as
\begin{equation}
\begin{aligned}
{\rm Score}_{\rm global}(v_i)
= \sum_{j=1}^{3}\Big( \mathrm{Count}\big(\mathrm{L}_j(v_i), \mathcal{A}_u\big) - 1 \Big),
\end{aligned}
\end{equation}
where $\mathrm{Count}(\mathrm{L}_j(v_i), \mathcal{A}_u)$ counts how many items in $\mathcal{A}_u$ share the same length-$j$ prefix with $v_i$, and the subtraction of $1$ removes the trivial self-count.

We rank all behaviors in $\mathcal{A}_u$ by ${\rm Score}_{\rm global}(v_i)$ and select the Top-$K$ items to form the global-aware subsequence:
\begin{equation}
\begin{aligned}
\mathcal{M}_u^{\rm global} &&= \mathrm{TopK}({\rm Score}_{\rm global}(v_i))
\end{aligned}
\end{equation}

\begin{figure}[!t]
\centering
\includegraphics[width=1.0\linewidth]{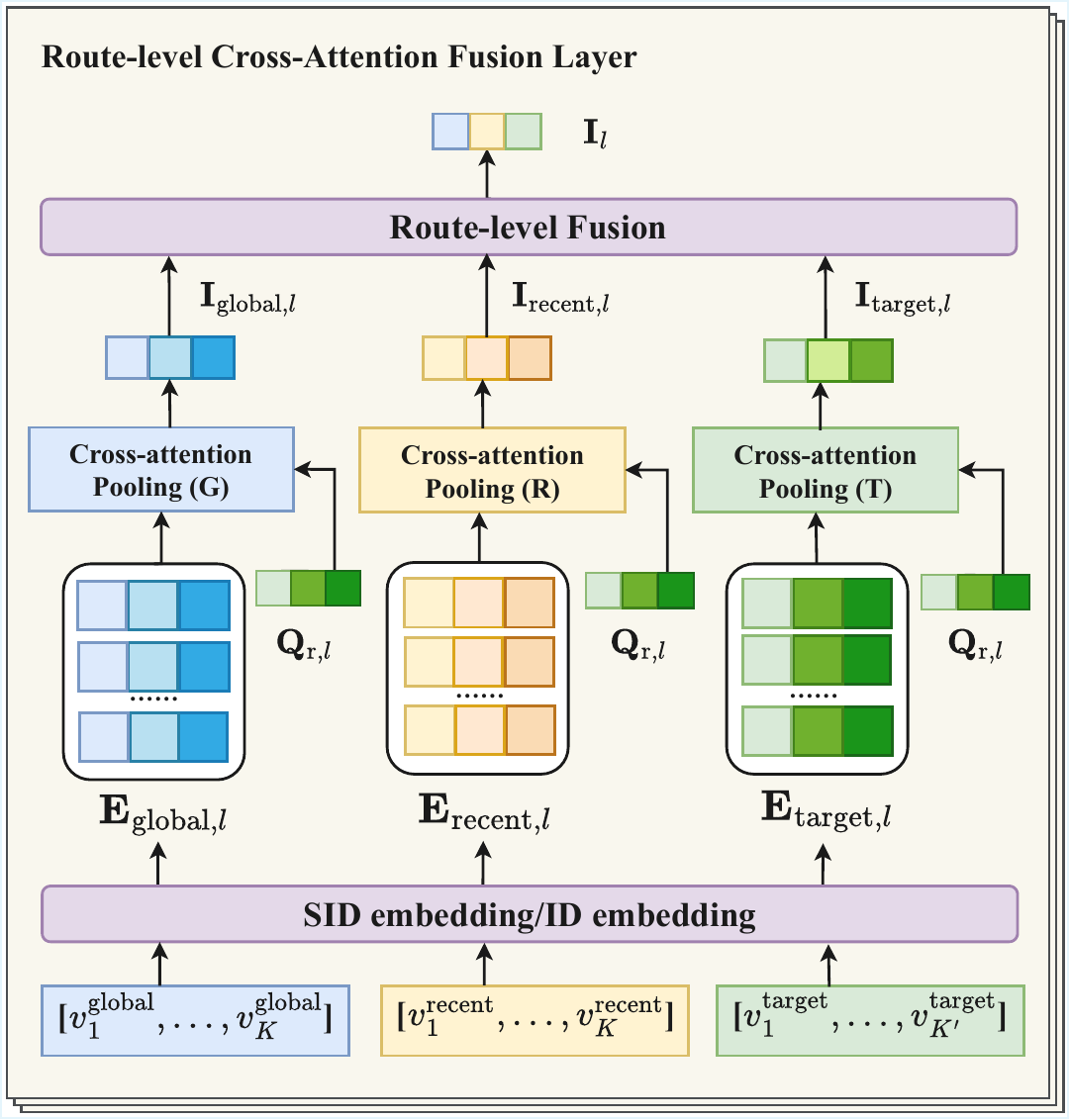}
\caption{The architecture of the Route-level Cross-Attention Fusion Layer. }
\label{fig:layer}
\vspace{-2mm}
\end{figure}

\subsection{Bi-level Fusion Refinement} 
\label{sec:modelling}

After multi-route retrieval, we obtain multiple semantically relevant behavior subsequences $\mathcal{M}_{u,v_t}=\{\mathcal{M}_{u,v_t}^{\rm target}, \mathcal{M}_{u}^{\rm recent}, \mathcal{M}_{u}^{\rm global}\}$
from different routes. Each of them reflects user interests from a distinct perspective and granularity.
Besides, the semantic representations encoded by hierarchical SIDs are not directly aligned with general ID-based features used for prediction.
To fuse interests at both route and SID levels, we propose a \textbf{Bi-level Fusion Refinement (BFR)}.
It consists of a \textbf{Route-level Cross-Attention Fusion Layer} and an \textbf{SID-level Gated Fusion}.
The former component is for alignment at the route level, while the latter one is at the SID level.

\subsubsection{\textbf{Route-level Cross-Attention Fusion Layer}}

As shown in Figure~\ref{fig:layer}, given the retrieved subsequences from three routes, we conduct target-aware interaction
to refine the information within each route.
For each route $r\in\{{\rm target},{\rm recent},{\rm global}\}$, the retrieval module returns
a subsequence $\mathcal{M}_{u}^{r}$.
At SID and ID level $l\in\{{\rm ID},{\rm SID}^{(1)},{\rm SID}^{(2)},{\rm SID}^{(3)}\}$,
we obtain the corresponding token representations by embedding lookup, \ie $\mathbf{E}_{r,l} = \mathbf{e}_l\!\left(\mathcal{M}_{u}^{r}\right)$, where $\mathbf{e}_l(\cdot)$ denotes the embedding function at level $l$.
Meanwhile, we denote the target representation at level $l$ as $\mathbf{Q}_{r,l}\in\mathbb{R}^{d}$.

To summarize each route in a target-aware manner, we adopt a cross-attention pooling mechanism.
Specifically, we project the target representation and the retrieved tokens into query, key, and value spaces:
\begin{equation}
\mathbf{Q} = \mathbf{Q}_{r,l}\mathbf{W}_Q,\qquad
\mathbf{K} = \mathbf{E}_{r,l}\mathbf{W}_K,\qquad
\mathbf{V} = \mathbf{E}_{r,l}\mathbf{W}_V,
\end{equation}
where $\mathbf{W}_Q,\mathbf{W}_K,\mathbf{W}_V\in\mathbb{R}^{d\times d}$ are learnable matrices.
Based on these projections, we derive a target-aware route representation by applying
cross-attention pooling over the retrieved tokens. Specifically, we compute attention weights
between the target query and each retrieved token, and aggregate the token representations
into a single route-level vector. To preserve the original target semantics, we further employ
a residual connection:
\begin{equation}
\mathbf{I}_{r,l}
=
\underbrace{\mathrm{softmax}\!\left(\frac{\mathbf{Q}\mathbf{K}^{\top}}{\sqrt{d}}\right)\mathbf{V}}_{\text{Cross-Attention Pooling}}
\;+\;
\mathbf{Q}_t^{\,l},
\qquad r\in\{{\rm target},{\rm recent},{\rm global}\}.
\label{eq:xattn}
\end{equation}
Here, the attention weights $\mathrm{softmax}\!\left(\frac{\mathbf{Q}\mathbf{K}^{\top}}{\sqrt{d}}\right)\in\mathbb{R}^{1\times K}$
assign higher attention weights to retrieved tokens that are more relevant to the target query, and the multiplication with $\mathbf{V}\in\mathbb{R}^{K\times d}$ performs a weighted sum over all retrieved tokens, thereby
pooling the subsequence into a single $d$-dimensional vector.

After obtaining the target-aware route representations
$\{\mathbf{I}_{r,l}\}_{r\in\{{\rm target},{\rm recent},{\rm global}\}}$ at level $l$,
we further fuse them to form a unified interest vector for this level.
Concretely, we compute route weights via a lightweight gating network and normalize the scores to obtain route weights:
\begin{equation}
\mathbf{s}_{l} = g_l(\mathbf{I}_{{\rm target},l}, \,\mathbf{I}_{{\rm recent},l},\,\mathbf{I}_{{\rm global},l}), \boldsymbol{\alpha}_{l}=\mathrm{softmax}(\mathbf{s}_{l}),
\end{equation}
where $g_l(\cdot)$ denotes a learnable function that outputs tri-route scores at level $l$.
The fused interest at level $l$ is then given by
\begin{equation}
\mathbf{I}_{l}=
\sum_{r\in\{{\rm target},{\rm recent},{\rm global}\}}
\alpha_{r,l}\,\mathbf{I}_{r,l}
.
\label{eq:route_gate}
\end{equation}

\subsubsection{\textbf{SID-level Gated Fusion}}
After route fusion, we obtain a level-specific interest representation $\mathbf{I}_{l}$ for traditional ID level and each SID level, \ie $l\in\{{\rm ID}, {\rm SID}^{(1)}, {\rm SID}^{(2)}, {\rm SID}^{(3)}\}$.
We further fuse these representations across levels to form a unified long-term interest vector of users.

Concretely, we compute level scores via a lightweight gating network and normalize the scores to obtain level weights:

\begin{equation}
\mathbf{s} = h\!\left(\mathbf{I}_{\bf{ID}},\,\mathbf{I}_{\bf{SID}^{(1)}},\,\mathbf{I}_{\bf{SID}^{(2)}},\,\mathbf{I}_{\bf{SID}^{(3)}}\right),
\boldsymbol{\beta}=\mathrm{softmax}(\mathbf{s}),
\end{equation}
where $h(\cdot)$ denotes a learnable function that outputs four-level scores.
Finally, the fused hierarchical representation is as follows:
\begin{equation}
\mathbf{I}_{u}=
\beta_{\bf{ID}}\mathbf{I}_{\bf{ID}}
+\beta_{1}\mathbf{I}_{\bf{SID}^{(1)}}
+\beta_{2}\mathbf{I}_{\bf{SID}^{(2)}}
+\beta_{3}\mathbf{I}_{\bf{SID}^{(3)}}
.
\label{eq:level_fuse}
\end{equation}

\begin{algorithm}[t]
\caption{Training and Inference of \modelname}
\label{alg:train}
\raggedright

\begin{algorithmic}[1]
    \State Specify semantic levels $\{{\rm SID}^{(1)},{\rm SID}^{(2)},{\rm SID}^{(3)}\}$.
    \State Specify subsequence length $K$ and recent window size $W$.
    \State Specify the modality encoder $E(\cdot)$ and SID tokenizer $\mathrm{Tokenizer}(\cdot)$.
    \State Generate multi-level SIDs for all items offline:
    $C_i\leftarrow \mathrm{Tokenizer}(E(s_i)),\ \forall v_i\in\mathcal{V}$.
\end{algorithmic}

\textbf{Train Process}
\setcounter{algorithm}{4}
\begin{algorithmic}[1]
    \makeatletter
    \setcounter{ALG@line}{4}
    \For{$(u,i,y)\in \mathcal{D}$}
        \State Encode the target item and user behaviors into token embeddings at all semantic levels.
        \State Perform tri-route retrieval to obtain subsequences
        $\{\mathcal{M}_u^{r}\}_{r\in\{{\rm target},{\rm recent},{\rm global}\}}$ (Sec.~\ref{sec:method_mmr}).
        \For{$l\in\{{\rm ID}, {\rm SID}^{(1)}, {\rm SID}^{(2)}, {\rm SID}^{(3)}\}$}
            \State Obtain token representations $\mathbf{E}_{r,l}$ via embedding lookup and index-based gathering.
            \State Compute route representations $\mathbf{I}_{r,l}$ using cross-attention pooling (Eq.~\eqref{eq:xattn}).
            \State Fuse routes to obtain $\mathbf{I}_{l}$ with the route gate (Eq.~\eqref{eq:route_gate}).
        \EndFor
        \State Fuse semantic levels to obtain the final long-term interest representation $\mathbf{I}_u$ (Eq.~\eqref{eq:level_fuse}).
        \State Predict $\hat{y}_{u,i}$ with the prediction layer.
        \State Compute the training loss and update model parameters by back-propagation.
    \EndFor
\end{algorithmic}

\textbf{Inference Process}
\setcounter{algorithm}{15}
\begin{algorithmic}[1]
    \makeatletter
    \setcounter{ALG@line}{16}
    \State Cache item SIDs and retrieval indices/statistics offline for efficient serving.
    \For{each query pair $(u,i)$}
        \State Run the same tri-route retrieval and hierarchical fusion procedure (without gradient) to obtain $\hat{y}_{u,i}$.
        \State Rank candidate items according to $\hat{y}_{u,i}$.
    \EndFor
\end{algorithmic}
\end{algorithm}

\subsection{Training and Inference}

We summarize the training and inference procedures of \modelname\ in Algorithm~\ref{alg:train}.
The model is trained by minimizing the binary cross-entropy loss:
\begin{equation}
\mathcal{L}_{\mathrm{ctr}}
=
-\frac{1}{\lvert \mathcal{D}\lvert}\sum_{(u,i,y)\in\mathcal{D}}
\Big(
y\log \hat{y}_{u,i} + (1-y)\log(1-\hat{y}_{u,i})
\Big).
\label{eq:loss}
\end{equation}

\noindent\textbf{Training.}
As shown in Algorithm~\ref{alg:train}, we first generate item SIDs offline (lines 1--4).
For each training instance (line 5), we encode the target item and user behaviors into token embeddings at all semantic levels (line 6), and perform tri-route retrieval to obtain subsequences (line 7).
For each SID and ID level (line 8), we gather route tokens by embedding lookup (line 9), compute route representations via cross-attention pooling (line 10), and fuse the three routes with the route gate to obtain $\mathbf{I}_{l}$ (line 11).
After iterating all levels, we fuse level-specific interests to obtain the final long-term representation $\mathbf{I}_u$ (line 13), predict $\hat{y}_{u,i}$ (line 14), and update model parameters by back-propagation with Eq.~\eqref{eq:loss} (line 15).

\vspace{1mm}
\noindent\textbf{Inference.}
In terms of inference, we cache item SIDs and retrieval statistics offline (line 17).
For each query pair $(u,i)$ (line 18), we run the same retrieval and hierarchical fusion pipeline (line 19) to obtain $\hat{y}_{u,i}$,
and rank candidate items accordingly (line 20).

%% file: 4Experiment.tex
\section{Experiment}





\subsection{Experimental Settings}
\subsubsection{\textbf{Datasets}}
\begin{table}[!t]
\centering
\caption{The statistics of the preprocessed datasets.}
\begin{tabular}{ccccc}
\toprule[1pt]
\textbf{Dataset} & \textbf{\# Users} & \textbf{\# Items} & \textbf{\# Interactions} &  \textbf{Avg.Length} \\ 
\midrule
JD & 34,546 &  592,214 & 3,356,860 & 97.17 \\
Pixel-1M & 168,776 & 99,016 & 8,168,918 & 48.40 \\
\bottomrule[1pt]
\end{tabular}
\label{tab:dataset}
\vspace{-2mm}
\end{table}
We apply two different domains and large-scale datasets to evaluate the effectiveness of our proposed \modelname, \ie JD\footnote{\url{https://github.com/guyulongcs/IJCAI2019_HGAT}}
 and Pixel\footnote{\url{https://github.com/westlake-repl/PixelRec}}. 
For the preprocessing, we first filter out the users who have fewer than $3$ interactions in JD and $28$ in Pixel.
Then, we adopt the same data preprocessing method as MIRRN~\cite{xu2025mirrn}.
All samples are divided into a training set (80\%), a validation set (10\%), and a test set (10\%) based on the click time of the target item.
For each user, we use their most recent 300 behaviors.
The statistical details of the preprocessed
two datasets are presented in Table \ref{tab:dataset}.

\subsubsection{\textbf{Baselines}}
To evaluate the effectiveness of \modelname, we compare it with several influential CTR prediction baselines, which are grouped into three categories.

\noindent The first category refers to the unified \textbf{Traditional Behavior Sequence Modeling} methods.
\begin{itemize}[leftmargin=*]
    \item \textbf{DIN}~\cite{zhou2018din}. DIN proposes a local activation unit to learn the representation of user interests from historical behaviors with respect to the target item.
    \item \textbf{DIEN}~\cite{zhou2019deep}. Following DIN, DIEN additionally adopts an interest extractor layer to capture temporal interests from the historical behavior sequence.
    \item \textbf{BST}~\cite{chen2019behavior}. It proposes to adopt the Transformer model~\cite{vaswani2017attention} to capture the sequential patterns underlying users' behavior sequences for recommendation.
    \item\textbf{TransAct}~\cite{xia2023transact}. TransAct focuses on short-term preferences by modeling users' real-time activities with a sequential model.

\end{itemize}
The second category includes the typical two-stage \textbf{Lifelong User Modeling} methods mentioned in Section~\ref{sec:pre_lifelong}
\begin{itemize}[leftmargin=*]
    \item \textbf{SIM}~\cite{pi2020sim}. It extracts user interests with two cascaded search units (GSU and ESU). The non-parametric SIM (hard) refers to using the category ID for searching, while SIM(soft) uses the dot product of pre-trained embeddings as the relevance metric.
    \item \textbf{ETA}~\cite{chen2021eta}.  It proposes a hash-based efficient target attention network to enable retrieval via low-cost bit-wise operations.
    \item \textbf{TWIN}~\cite{chang2023twin}. Its Retrieval stage adopts the identical target-behavior relevance metric as the target attention in the Refinement stage, making the two stages twins.
    
    \item \textbf{SDIM}~\cite{cao2022sampling}. It hashes the candidate and history items and gathers behaviors with matching signatures.

    \item \textbf{MIRRN}~\cite{xu2025mirrn}. MIRRN retrieves multi-granularity behavior subsequences, refines them with a Fourier transformer, and fuses them via target-aware attention for CTR prediction.

\end{itemize}
Furthermore, we also include baselines of semantic utilization.
\begin{itemize}[leftmargin=*]
    \item \textbf{ReLLa}~\cite{lin2024rella}. It designs a semantic retrieval module to improve the data quality of data samples and then prompts LLMs for CTR prediction. For the implementation in this paper, we adopt the few-shot fine-tuning version.
    \item \textbf{SEmb}. It refers to a variation of our \modelname~ that replaces SID with semantic embeddings directly.
\end{itemize}

\subsubsection{\textbf{Implementation Details}}
We implement \modelname~and all baselines on a machine equipped with one NVIDIA RTX 4090D GPU (24GB) and 15 vCPUs (Intel Xeon Platinum 8474C). The required software environment includes CUDA 12.4, Python 3.12.3, and PyTorch 2.6.0. 
For SID generation, we use RQ-KMeans~\cite{deng2025onerec} as the tokenizer and Flan-T5-XL~\cite{chung2024scaling} as the encoder. 
Following GRID~\cite{grid}, we set the number of residual layers to $L=3$ and the number of tokens per layer to $D=256$.
For a fair comparison, our model and all baselines share the same overall network structure and input features. In addition, we use the same Top-$K$ setting
for behavior retrieval across all retrieval-based methods, where $K$ is set to $20$.
Besides, in recent route, $W$ and $N$ are fixed to $16$ and $2$, respectively.
For all methods, the prediction head is an MLP with layer sizes $[200, 80, 2]$.
We set the embedding dimension to 16 for all feature fields and use a batch size of 256.
All models are trained with Adam using a fixed learning rate of 0.001 for one epoch.
We report the average values across $3$ runs with the same random seed set.

\subsubsection{\textbf{Evaluation Metrics}}
Following previous works~\cite{zhu2021open}, we adopt two most commonly-used metrics, \ie AUC and logloss.

\begin{itemize}[leftmargin=*, itemsep=0pt, topsep=2pt]
  \item \textbf{AUC} (Area Under the ROC Curve) measures the probability that a randomly selected positive sample is ranked higher than a randomly selected negative sample. A higher AUC indicates better CTR prediction performance.
  \item \textbf{Logloss}, which is known as binary cross-entropy loss, is defined in Eq.~\eqref{eq:loss}. A lower logloss indicates better performance.
\end{itemize}

\begin{table}[!t]
\centering
\caption{Overall performance of various methods on two public datasets. The boldface refers to the highest score and the underline indicates the best result of the methods. ``\textbf{{\Large *}}'' indicates the statistically significant improvements (\ie two-sided t-test with $p<0.05$) over the best baseline.}
\resizebox{1.0\columnwidth}{!}{
\begin{tabular}{c|cc|cc}
\toprule[1pt]
\multirow{2}{*}{\textbf{Model}} 
& \multicolumn{2}{c|}{\textbf{JD}} 
& \multicolumn{2}{c}{\textbf{Pixel-1M}} \\ \cmidrule{2-5}

& \textbf{AUC} & \textbf{LogLoss} 
& \textbf{AUC} & \textbf{LogLoss} \\

\midrule

DIN~\cite{zhou2018din}       &0.7509  &0.5802  &0.6632  &\ul{0.6805}  \\
DIEN~\cite{zhou2019deep}       &0.7371  &0.5878  &0.6718  & 0.7123 \\
BST~\cite{chen2019behavior}       &0.7281  &0.6154  &0.6646  &0.7440  \\
TransAct~\cite{xia2023transact}  &0.7470  &0.5776  &0.6658  &0.7367  \\
\midrule
SIM (soft)~\cite{pi2020sim} &0.7746  &0.5560  &0.6635  &0.7279  \\
SIM (hard)~\cite{pi2020sim}     &\ul{0.7944}  &\ul{0.5425}  &0.6601  &0.7093  \\
ETA~\cite{chen2021eta}       &0.7641  &0.5733  &0.6605  &0.7314  \\
TWIN~\cite{chang2023twin}      &0.7776  &0.5540  &0.6607  &0.7086  \\
SDIM~\cite{cao2022sampling}      &0.7694  &0.5639  &0.6605  &0.7360  \\
MIRRN~\cite{xu2025mirrn}     &0.7602  &0.5726  &0.6653  &0.7149  \\
\midrule
ReLLa~\cite{lin2024rella} &0.6771
 &0.6749
 &0.6160
 &\textbf{0.6752}
 \\
SEmb &0.7849 &0.5480 &\ul{0.6721} &0.7157 \\
\midrule
\cellcolor{blue!10}\textbf{\modelname}       &\cellcolor{blue!10}\textbf{0.8117}*  &\cellcolor{blue!10}\textbf{0.5210}*  &\cellcolor{blue!10}\textbf{0.6729}*  &\cellcolor{blue!10}0.7142  \\
\bottomrule[1pt]
\end{tabular}
}
\label{tab:jd_pixel_results}
\vspace{-2mm}
\end{table}

\subsection{Overall Performance}
In Table~\ref{tab:jd_pixel_results}, we show the overall performance of our \modelname ~and all the baselines on two public datasets. In general, our model achieves a higher AUC than all traditional behavior sequence modeling methods and all Lifelong user modeling methods on both datasets. Its LogLoss results are also generally among the best, indicating the effectiveness of the proposed approach for CTR prediction with long user histories. We next provide a more detailed analysis.

Among the traditional behavior sequence modeling methods, DIN and DIEN achieve the best performance on both datasets, indicating that target-aware interest modeling is highly effective. DIN activates user interests conditioned on the target item, while DIEN further captures the evolution of such interests along the behavior sequence. Their strong results align with our motivation that accurately identifying and aggregating target-relevant interests is crucial, especially when users exhibit diverse preferences and interest drift over time. Building on this insight, our \modelname ~further improves performance by modeling users' preferences at various time granularities and reducing the impact of noisy or weakly relevant interactions. In this way, \modelname~provides a more robust and comprehensive user representation than DIN and DIEN.

Among the lifelong user modeling methods, SIM and MIRRN achieve the strongest overall performance on the two datasets.
However, SIM and MIRRN mainly rely on collaborative signals derived from user-item interactions when selecting and aggregating behaviors. In contrast, our \modelname~ further strengthens long-term modeling by introducing semantic information to better identify and aggregate target-relevant behaviors. 
By augmenting collaborative signals with semantic information, our approach produces a more robust long-term representation, leading to consistent gains over these strong long-sequence baselines. 

\modelname~also outperforms ReLLa and SEmb overall, highlighting its advantage in capturing target-relevant semantics and modeling users' long-term preferences. In particular, the gains over SEmb suggest that strict SID prefix matching is more effective at capturing fine-grained semantic interests. Moreover, the improvements over ReLLa indicate that jointly leveraging semantic and collaborative signals is beneficial for long-term preference modeling.
The lower logloss of ReLLa on the Pixel dataset may stem from LLMs' stronger memorization~\cite {tirumala2022memorization}, but fine-tuning LLMs is extremely costly.

\subsection{Ablation Study}
\begin{table}[!t]
\centering
\caption{Ablation study on components.}
\resizebox{1.0\columnwidth}{!}{
    \begin{tabular}{l|cc|cc}
        \toprule[1pt]
        
        \multirow{2}{*}{\textbf{Method}} 
        & \multicolumn{2}{c|}{\textbf{JD}} 
        & \multicolumn{2}{c}{\textbf{Pixel-1M}} \\ 
        \cline{2-5}
        
        & \textbf{AUC} & \textbf{LogLoss} & \textbf{AUC} & \textbf{LogLoss} \\
        \midrule
        
        \textit{w/o} SID Retrival     & 0.7715 & 0.5574 & 0.6682 & 0.7279 \\
        \textit{w/o} Tree-based     & 0.7883 & 0.5404 & 0.6711& 0.7296\\
        \textit{w/o} SID Feature    & 0.7662 & 0.5620 & 0.6657 & 0.7289 \\
        \textit{w} Avg     & 0.7637 & 0.5774 & 0.6693 & \textbf{0.6957} \\
        \textit{w} Self-Attn & 0.7893 & 0.5483 & 0.6671 & 0.7237 \\
        
        \midrule
        \cellcolor{blue!10}\textbf{\modelname} 
        & \cellcolor{blue!10}\textbf{0.8117} 
        & \cellcolor{blue!10}\textbf{0.5210} 
        & \cellcolor{blue!10}\textbf{0.6729} 
        & \cellcolor{blue!10}0.7142 \\
        \bottomrule[1pt]
    \end{tabular}
}
\label{tab:components}
\vspace{-1mm}
\end{table}

\begin{table}[!t]
\centering
\caption{Ablation study of retrieval routes on Pixel dataset.}
\resizebox{1.0\columnwidth}{!}{
    \setlength{\tabcolsep}{14pt}
    
    \begin{tabular}{ccc|cc}
        \toprule[1pt]
        \textbf{T} & \textbf{R} & \textbf{G} & \textbf{AUC} & \textbf{LogLoss} \\
        \midrule
        
        \checkmark & & & 0.6722 & 0.7145 \\
        & \checkmark & & 0.6724 & 0.7193 \\
        & & \checkmark & 0.6657 & 0.7322 \\
        
        \checkmark & \checkmark & & 0.6717 & 0.7122 \\
        \checkmark & & \checkmark & 0.6724 & 0.7245 \\
        & \checkmark & \checkmark & 0.6708 & 0.7146 \\
        
        \midrule
        
        \cellcolor{blue!10}\textbf{\checkmark} 
        & \cellcolor{blue!10}\textbf{\checkmark} 
        & \cellcolor{blue!10}\textbf{\checkmark} 
        & \cellcolor{blue!10}\textbf{0.6729} 
        & \cellcolor{blue!10}\textbf{0.7142} \\
        \bottomrule[1pt]
    \end{tabular}
}
\label{tab:routes}
\vspace{-3mm}
\end{table}
We conduct an ablation study to examine the contribution of each component in \modelname, reporting the results in Table~\ref{tab:components}.

To validate the effectiveness of our retrieval module, we first remove the SID-based retrieval and retrieve tri-route subsequences using traditional ID, denoted as \textbf{\textit{w/o} SID Retrieval}. This variant shows a clear performance drop, demonstrating that incorporating semantic information improves retrieval quality.
\textbf{\textit{w/o} Tree-based} further removes the tree-based matching mechanism,\ie scoring behaviors by the number of matched SIDs directly. The degraded performance indicates that the proposed prefix-tree retrieval can identify semantically relevant behaviors constructed by RQ-KMeans more accurately.

Then, to examine the effectiveness of fusing semantic and collaborative interests, \textbf{\textit{w/o} SID Feature} keeps only the ID-level cross-attention branch. The performance drop suggests that introducing semantic features complements collaborative signals and yields a richer representation of user interest.

To further validate the effectiveness of cross-attention, we conduct two additional ablations: \textbf{\textit{w} Avg} and \textbf{\textit{w} Self-attn}. The former replaces cross-attention with simple average pooling over the retrieved subsequence, while the latter uses self-attention without conditioning on the target item. Both variants lead to inferior performance, indicating that target-aware cross-attention is crucial for accurately capturing target-related signals at each level. 

To validate the effectiveness of the proposed three-route retrieval, we further conduct ablations by enabling different route combinations. The results are shown in Table~\ref{tab:routes}. We observe that using all three routes achieves the best performance, suggesting that the multi-route design can effectively capture user interests at different granularities, which leads to superior overall results.

\subsection{Efficiency Study}
\begin{figure}[!t]
\centering
\includegraphics[width=0.8\linewidth]{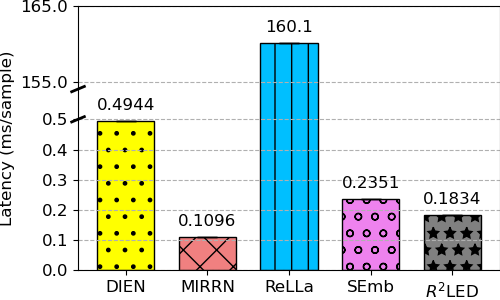}
\caption{Efficiency comparison (inference time per sample) between \modelname~ and baselines on the JD dataset.}
\label{fig:efficiency}
\vspace{-3mm}
\end{figure}
We evaluate the practical efficiency of \modelname~ by comparing its inference time per sample against representative baselines, including  DIEN, MIRRN, ReLLa, SEmb and \modelname.

As shown in Figure~\ref{fig:efficiency}, ReLLa is undoubtedly the most time-consuming, since it adopts LLMs for inference directly. 
DIEN has higher latency than MIRRN, SEmb and \modelname, indicating the efficiency of the two-stage framework.
SEmb is slower since it requires complex semantic embedding retrieval during inference.  
Though \modelname~ is less efficient than MIRRN due to the additional semantic modeling, it can achieve higher prediction accuracy.
Our \modelname~ can get a better trade-off between effectiveness and efficiency.

\subsection{Hyperparameter Analysis}
%

\begin{figure}[t]
    \centering
    \begin{subfigure}[t]{0.48\linewidth}
        \centering
        \includegraphics[width=\linewidth]{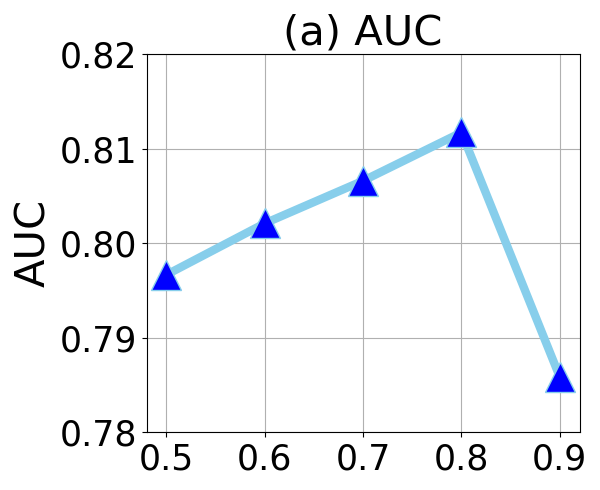}
        \label{fig:hp_auc}
    \end{subfigure}
    \hfill
    \begin{subfigure}[t]{0.48\linewidth}
        \centering
        \includegraphics[width=\linewidth]{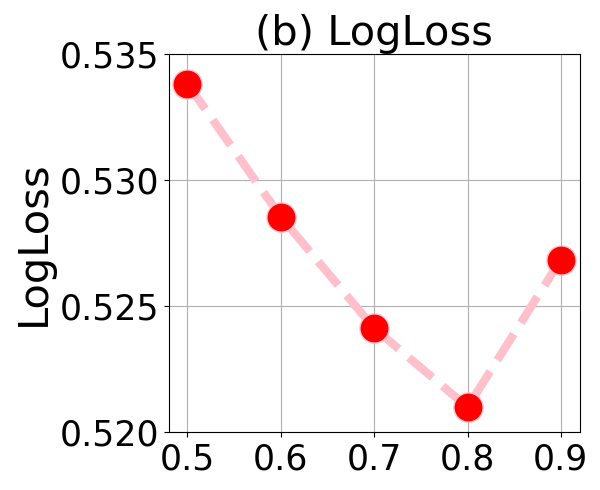}
        \label{fig:hp_logloss}
    \end{subfigure}

    \caption{Hyperparameter sensitivity analysis of \modelname~on the JD dataset.}
    \label{fig:hyperparam_2x2}
\vspace{-4mm}
\end{figure}
We vary the threshold $\tau$ that controls the confidence gate in Eq. \eqref{eq:fill_gate} and show the performance trends in Figure~\ref{fig:hyperparam_2x2}. Observing the results of $\tau$, we find that the model performance presents rising first and dropping then within the range of $[0.5, 0.9]$. Since $\tau$ only affects the retrieval filling module, this trend indicates that retrieval quality itself follows a similar pattern: when $\tau$ is too low, the gate is easily activated and introduces excessive noisy collaborative signals, while an overly large $\tau$ makes the gate too strict, so it often blocks filling and misses helpful collaborative signals.

\subsection{Long-tail Analysis}

To validate the effectiveness of introducing semantics, we conduct a long-tail analysis on the Pixel dataset, as shown in Table~\ref{tab:longtail}. Following the Pareto principle~\cite{box1986analysis}, we rank items by popularity and define the bottom 80\% as the Tail group, with the remaining items forming the Head group. 

Specifically, on the Head group, \modelname~achieves the best AUC and the lowest LogLoss, indicating more accurate ranking and probability estimation for popular items. More importantly, \modelname~also achieves a better performance on the Tail group, indicating that our method alleviates the semantic deficiency by leveraging SID effectively. 
In particular, our method generally outperforms SEmb, demonstrating that the use of SID can alleviate the single-granularity dilemma in semantic embedding.
Though ReLLa outperforms \modelname~on tail group since it utilizes semantic information by LLMs directly, it struggles to be deployed into real-time systems.

\begin{table}[!t]
\centering
\caption{Long-tail analysis on the Pixel dataset.}
\resizebox{0.85\columnwidth}{!}{
    \begin{tabular}{l|cc|cc}
        \toprule[1pt]
        
        \multirow{2}{*}{\textbf{Model}} 
        & \multicolumn{2}{c|}{\textbf{Head}} 
        & \multicolumn{2}{c}{\textbf{Tail}} \\ 
        \cline{2-5}
        
        & \textbf{AUC} & \textbf{LogLoss} & \textbf{AUC} & \textbf{LogLoss} \\
        \midrule
        ETA        & 0.6982 & 0.6148 & 0.5985 & 0.7660 \\
        MIRRN      & 0.6990 & 0.6103 & 0.6030 & 0.7560 \\
        ReLLa      & 0.6081 & 0.6510 & \textbf{0.6142} & \textbf{0.6764} \\
        SEmb       & 0.6946 & 0.5715&  0.6076 & \ul{0.7339} \\

        \midrule
        \cellcolor{blue!10}\textbf{\modelname} 
        & \cellcolor{blue!10}\textbf{0.6995} 
        & \cellcolor{blue!10}\textbf{0.5853} 
        & \cellcolor{blue!10}\ul{0.6132} 
        & \cellcolor{blue!10}0.7553 \\
        \bottomrule[1pt]
    \end{tabular}
}
\label{tab:longtail}
\vspace{-2mm}
\end{table}

\subsection{Case Study}
To compare SID-based retrieval with LSH-based retrieval, a case study is conducted on a representative user (ID: 87081) from the JD dataset, as illustrated in Figure~\ref{fig:case_study}. Based on the user’s most recent 300 behaviors, the history is dominated by mother-and-baby items with only a small portion of electronics. When the target item is ``iPhone XS (Space Gray)'', LSH-based retrieval is heavily biased by the dominant collaborative signals in the sequence, and thus mainly returns irrelevant mother-and-baby products. In contrast, SID-based retrieval can precisely identify semantically relevant behaviors and retrieve a diverse set of iPhone-related items, yielding a much more target-consistent subsequence.

%% file: 5RelatedWork.tex
\section{Related Works}

\subsection{Lifelong User Behavior Modeling}
User behavior modeling aims to learn users' preferences from their historical interactions. Early studies mainly focus on sequential behavior modeling with short- to medium-length histories. 
BST~\cite{chen2019behavior} adopts a Transformer-based architecture to capture the sequential signals in users' behavior sequences.
DIN~\cite{zhou2018din} introduces an attention mechanism to highlight and aggregate interests that are most relevant to the target item. 
DIEN~\cite{zhou2019deep} uses GRU with an update gate to strengthen the effect of relevant interests on the target item.

Given the extensive length of user behavior sequences, early works often struggle in practice due to the noise in long histories and the high computational cost of modeling ultra-long sequences. To address these challenges, subsequent studies have developed more effective architectures for user behavior modeling.
SIM~\cite{pi2020sim} adopts two cascaded search units with hard/soft retrieval to select target-relevant behaviors from ultra-long histories, thereby mitigating noise in lifelong recommendation.
ETA~\cite{chen2021eta} proposes an end-to-end model that leverages locality-sensitive hashing (LSH) to hash item embeddings and retrieve relevant items based on Hamming distance.
TWIN~\cite{chang2023twin} addresses the inconsistency in two-stage lifelong behavior modeling by using an identical target--behavior relevance metric in both stages.
Building on TWIN, TWIN-v2~\cite{si2024twinv2} further improves the efficiency of long-sequence modeling by introducing hierarchical clustering and cluster-aware target attention.
MIRRN~\cite{xu2025mirrn} retrieves subsequences over multiple time scales, enabling it to capture user interests at multiple granularities. 
However, existing works mainly focus on collaborative signals, facing the challenges of noisy sensitivity and semantic deficiency. This paper derives a SID-based framework to handle these issues.

\subsection{SID-enhanced Recommendation}
Recent advancements in large language models~\cite{wu2024survey,zhao2023survey} have revolutionized recommender systems (RS) into a new semantic paradigm. Some research studies~\cite{liu2024llm} have adopted semantic embeddings derived from LLMs to enhance RS.
However, the utilization of semantic embedding faces the challenges of single-granularity and inefficiency.
To address these constraints, SID-enhanced RS is proposed. It adopts quantization methods, \eg RQ-VAE~\cite{lee2022autoregressive}, to decompose continuous semantic features into sequences of discrete codes. Then, they are considered as a semantic feature fed into RS models.
For example, VQRec~\cite{VQRec}, SPM-SID~\cite{SPM_SID}, and PG-SID~\cite{prefixgram_SID} exploit the shared code indices of semantically akin items, a characteristic of SIDs. By consolidating interaction signals based on semantic proximity rather than potentially unreliable co-occurrence patterns, such techniques can establish semantic relationships between users and items. 
Concurrently, other frameworks like CCFRec~\cite{CCFRec} and PSRQ~\cite{PSRQ} harness the inherently hierarchical nature of SIDs to model distinct semantic units, while H2Rec~\cite{liu2025best} aims to fuse SID and hash ID more effectively.
Despite their progress, we are the first to explore the utilization of SID in lifelong user behavior modeling, facing the retrieval complexity and space misalignment challenges.

\begin{figure}[!t]
\centering
\includegraphics[width=1\linewidth]{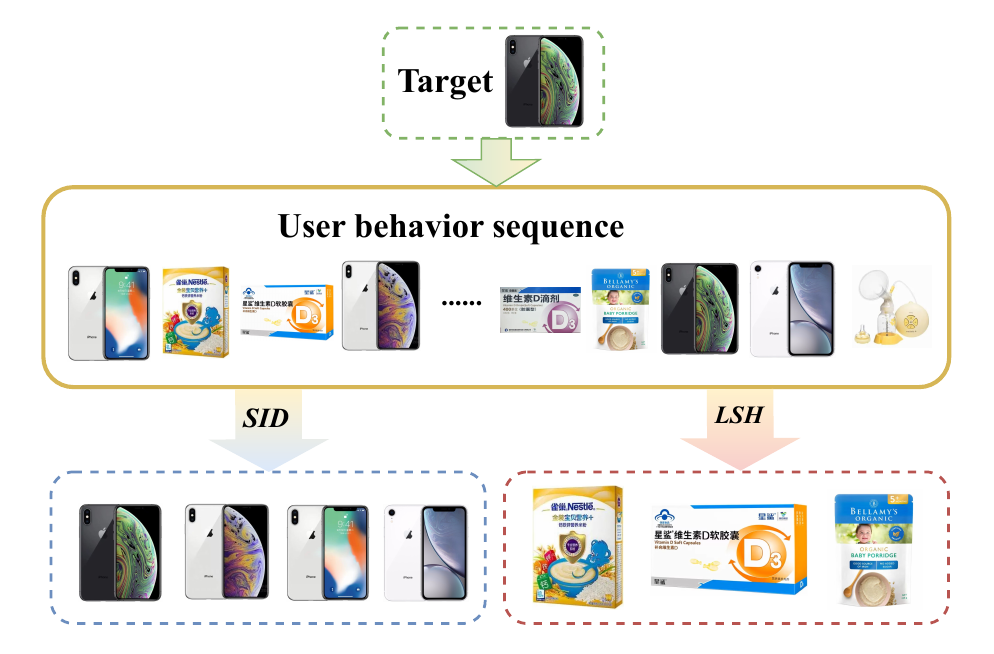}
\caption{The case study of user 87081 on the JD dataset.}
\label{fig:case_study}
\vspace{-4mm}
\end{figure}

%% file: 6Conclusion.tex
\section{Conclusion}

In this paper, we propose \modelname~ for lifelong user behavior modeling in CTR prediction. To mitigate noise in retrieval and overcome limited semantic understanding during refinement, \modelname~ adopts a semantic-aware retrieval-and-refinement framework that better identifies and aggregates target-relevant long-term behaviors. Specifically, it leverages items' semantic IDs to retrieve multi-granularity behavior subsequences from users' ultra-long histories, and then performs bi-level cross-attention fusion to combine these semantic interests into a unified target-aware representation. The effectiveness and efficiency of \modelname~ is validated through experiments on two public datasets. 
